\documentclass[12pt]{article}
\usepackage{epsfig}

\def\beq{\begin{equation}}
\def\eeq{\end{equation}}
\def\bea{\begin{eqnarray}}
\def\eea{\end{eqnarray}}
\def\ksl{\hbox{\hbox{${k}$}}\kern-1.9mm{\hbox{${/}$}}}

\def\dofiga#1#2{\centerline{
\epsfxsize=#1\epsfig{file=#2, width=13cm,height=11.0cm, angle=0}
\hspace{0cm}
}}
\def\dofigb#1#2{\centerline{
\epsfxsize=#1\epsfig{file=#2, width=10cm,height=8cm, angle=0}
\hspace{0cm}
}}

\newcommand{\nn}{\nonumber}
\newcommand{\text}{\rm}

\newcommand{\Mp}{\bar{M}_P}

\def\lsim{\raise0.3ex\hbox{$\;<$\kern-0.75em\raise-1.1ex\hbox{$\sim\;$}}} 

\def\gsim{\raise0.3ex\hbox{$\;>$\kern-0.75em\raise-1.1ex\hbox{$\sim\;$}}}

\begin{document}
\begin{titlepage}

{\flushright{
        \begin{minipage}{5cm}
          CERN-PH-TH/2008-246\\
          RM3-TH/08-22
        \end{minipage}        }

}
\vskip 2cm
\begin{center} 

\boldmath
{\Large\bf 
Flavor Changing Fermion-Graviton Vertices}

\vspace{.3cm}

G. Degrassi$^{a,b}$, E. Gabrielli$^b$, and L. Trentadue$^c$
\vskip 1.cm

\emph{
$^a$ Dipartimento di Fisica, Universit\`a di Roma Tre and INFN, Sezione
di Roma Tre,
Via della Vasca Navale 84, I-00146 Rome, Italy\\
$^b$ CERN PH-TH, CH-1211 Geneva 23, Switzerland\\
$^c$ Dipartimento di Fisica, Universit\`a di Parma
and INFN, Gruppo Collegato di Parma,
Viale delle Scienze 7, I-43100 Parma, 
Italy
}
\end{center}

\vskip 0.7cm
\begin{abstract}
We study the flavor-changing quark-graviton vertex
that is induced at the one-loop level when gravitational
interactions are coupled to the standard model.
Because of the conservation of the energy-momentum tensor 
the corresponding form factors turn out to be finite and gauge-invariant. 
Analytical expressions of the form factors
are provided at leading order in the external masses.
We show that flavor-changing interactions in gravity are local
if the graviton is strictly massless while if the graviton has a small mass
long-range interactions inducing a flavor-changing contribution in 
the Newton potential appear. Flavor-changing processes with massive spin-2
particles are also briefly discussed. These results can be generalized to 
the case of the lepton-graviton coupling.

\end{abstract}
\vfill
\end{titlepage}

\section{Introduction}
Gravity is the only fundamental interaction which is universally coupled to 
all matter fields. 
Universality is then guaranteed by the tree-level coupling of the 
graviton field with the conserved energy-momentum tensor.
However, when quantum corrections are taken into account, matter-coupling
universality could be spoiled.

It is known that quantum corrections to 
the graviton-matter vertices (without gravitons in the loops)
are finite if the underlying theory of matter fields
(in flat space-time) is renormalizable 
\cite{Delb,Grisaru,BG1,kob-okun,milton,BG2}. 
This remarkable property is just a 
consequence of the Ward Identities (WI) that result from
the energy-momentum conservation \cite{BG1}. 
Therefore, (matter) radiative corrections
to the graviton-matter vertices can be consistently calculated 
in the framework of gravity coupled to the standard model (SM) theory.

Long ago, Berends and Gastmans evaluated
the finite corrections to the graviton-photon and graviton-electron 
vertices due to virtual quantum electrodynamical (QED) 
processes at the one-loop 
level\footnote{The same calculation was previously done
by Delbourgo and Phocas-Cosmetatos in \cite{Delb}, 
but it was incorrect due to the wrong Feynman rules adopted \cite{BG2}.} 
\cite{BG2}. In the latter case the corrections found
lead to a modification of the Newton's law.
In particular, for an electron (or any charged particle) in a 
gravitational field of a stationary mass $M$, 
they found a repulsive correction term  proportional
to the classical radius of the  particle, $r_e = \alpha/m_e$, or
\bea
V(r)\sim -\frac{G_N M m_e}{r}\left(1-\frac{\alpha}{2 m_e r}\right)\, ,
\label{NP}
\eea
where $G_N$ is the Newton constant, $\alpha$ is the fine-structure
constant, and $m_e$ is the mass of the electron. At macroscopic distances  
($r\gg r_e$) the modification of 
Newton's law induced by the $1/r^2$ term in Eq.(\ref{NP}) is extremely
small, ${\cal O}(10^{-22})$ at the surface of the earth \cite{BG2}.

The term $1/r^2$ has an infrared origin as it can be simply 
understood by looking at the singular behavior of the one-loop 
Feynman diagrams in the massless limit ($m_e\to 0$).
The correction to the tree-level electron-graviton vertex 
is proportional to  $\alpha \sqrt{-q^2/m_e^2}$ 
for negative small values of $q^2$, where $q$ is the momentum transfer.
This is due to the fact that the photon has a tree-level
coupling with the graviton.  Then,  the diagram where
the external graviton is attached to two virtual photons
in the loop contains a term with  a powerlike infrared singularity
as $m_e\to 0$.
Thus, after Fourier transforming  into coordinates space, 
this contribution gives rise to a  $1/(r^2m_e)$ correction to the Newton 
potential.
Notice that there is no counterpart of such effect in QED.
Indeed, due to the absence of self-photon interactions at tree-level,
the Coulomb potential is protected against $1/r^2$ corrections
of order ${\cal O}(\alpha)$.

It is remarkable that, 
although derived in the framework of quantum field theory, 
the $1/r^2$ term in Eq.(\ref{NP}) has a pure classical origin and it
was first obtained on the basis of purely classical considerations in 
Ref.\cite{DeWitt}. Indeed this term 
arises from the fact that the total mass of the particle is not
concentrated at a point but is partly distributed as field energy in the space
around the particle \cite{DeWitt}.

When gravity is coupled to the weak interactions the one-loop radiative 
corrections  to the graviton-fermion 
vertex will include also the virtual exchanges of the 
weak gauge bosons $W^{\pm}$ and $Z$.
At the moment, a complete study of these effects is still missing. 
These corrections are expected to give small contributions 
in the low energy limit and  furthermore to be infrared safe, being the 
$W^{\pm}$ and $Z$ massive. Therefore, no long-distance modifications 
in the Newton potential are expected from the weak radiative corrections. 

A peculiar aspect of the weak interactions
is that they can induce (at one loop) flavor-changing 
neutral currents (FCNC) processes in the fermionic sector. Since
FCNC are absent at tree-level, they arise as a pure 
quantum effect. Analogously, flavor-changing (FC) graviton vertices are not 
present at the tree-level.
However, when gravity is coupled to the weak
interactions,  quantum weak-corrections can induce 
an off-diagonal  contribution (in flavor space) to the energy-momentum tensor. 
These effects are the spin-2 counterpart of the standard spin-1 FCNC 
contribution in the SM. Because of the tensorial nature of the coupling of 
spin-2 particles to the matter fields we will refer to these effects as 
tensorial flavor-changing neutral currents (TFCNC).

The  aim of this paper is to compute the
form factors of the TFCNC and analyze their gravitational couplings.
We  perform the exact computation of the one-loop FC fermion-graviton
vertex, i.e.~retaining the full  dependence on all masses and momenta.
We investigate under which conditions these effective couplings
could induce  modifications 
in the Newton potential discussing both cases of massless and 
massive-graviton exchange. We consider a few  applications of these
results in the case of massive spin-2 particles.
In particular, we analyze the FC decay $f_1\to f_2 G$, 
and the  decay $G\to f_1 \bar{f}_2$  where $G$ stands for a 
massive spin-2 particle. Our results
can be easily generalized to the leptonic sector with massive 
Dirac neutrinos, provided the CKM matrix of the quark sector 
is substituted with the corresponding leptonic one.

The paper is organized as follows. In section 2 we specify the interaction
between the SM fields and the graviton by suitably choosing a gauge
that avoids the appearance of cubic vector boson-Goldstone-graviton
interactions. We analyze the 
general structure of the off-diagonal matrix element (in flavor space) 
of the energy-momentum tensor of fermion fields, 
derive the corresponding Ward-identities, and provide the analytical 
expressions of the relevant form factors. 
The analysis of the gravitational couplings 
of the TFCNC is carried out in Sec. 3.
In Sec. 4 we  analyze a few  applications of these
results. Finally, Sec. 5 contains our conclusions.

\section{Flavor-changing quark-graviton vertex}
We are going  to
evaluate  the interaction between  the graviton
and two generic fermion states with different flavor or  
the off-diagonal
matrix element of the energy-momentum tensor $T^{\mu\nu}(x)$.
In particular, working in the basis of mass-eigenstates 
for the fermion fields, we are interested in calculating the following matrix
element\footnote{
The definition of $\hat{T}^{\mu\nu}$ corresponds to 
the Feynman rule for the off-diagonal matrix element of 
the energy-momentum tensor.}
\beq
\hat{T}^{\mu\nu}\equiv -i\langle p_2 |~ T^{\mu\nu}(0)~ | p_1\rangle\, ,
\label{eq:1}
\eeq
where the  initial and final states are assumed to have momenta 
$p_1$ and $p_2$ respectively with $p_i^2 = m_i^2$  and
spinorial wave-functions $u_i(p_i)$ $(i=1,2)$ while the 
four-momenta $p,~ q$ are defined as $p=p_1+p_2$ and $q=p_1-p_2$.
It is understood that initial and final states
have different flavor and mass. The 
relevant Feynman rules which are necessary in
order to calculate the above matrix element, can be easily derived by looking
at the graviton couplings with matter fields.

The interactions between gravity and the SM fields are assumed to be described
by the action integral
\beq
{\cal S} =\int d^4 x \, \sqrt{- \rm g}\, {\cal L}_{\rm SM} [g_{\mu\nu}]
\label{eq:2}
\eeq
where the SM Lagrangian, ${\cal L}_{\rm SM}$, is thought to be written in 
terms of the metric tensor
$g_{\mu \nu}$ and for the fermionic part using the vierbein formalism.
In Eq.(\ref{eq:2}) ${\rm g} = det\, g_{\mu\nu}$ and 
${\cal L}_{\rm SM}$
includes the classical term  and the gauge-fixing function in the 
$R_\xi$ gauge (the term with the ghost fields is not relevant for our 
discussion).
To obtain the interaction of the graviton with the SM fields 
we expand in Eq.(\ref{eq:2}) the metric $g_{\mu\nu}$ around the flat one as 
$g_{\mu\nu}= \eta_{\mu\nu} + \kappa h_{\mu\nu}$, retaining only the first
term in the graviton field  where 
$\eta_{\mu\nu} =(+1,-1,-1,-1)$, $\kappa = \sqrt{32\pi G_N}$, 
and $h_{\mu\nu}$  is interpreted as the spin-2 graviton field.
We notice that, if in the gauge-fixing term in ${\cal L}_{\rm SM}$ the
standard 't-Hooft gauge-fixing function is taken, vertices in which a graviton
field induces a transition between a vector boson and its unphysical 
Goldstone boson counterpart are going to appear \cite{CPRS}. In order 
to avoid the appearance of  vector-Goldstone-graviton interactions we find it 
convenient to use a modified version of the 't-Hooft gauge-fixing function, 
or
\bea
{\cal L}_{g.f.} & = &- \frac1{2 \xi} \left[
g_{\mu\nu} \partial^\mu W_a^\nu +g^{\nu\rho} \Gamma^\mu_{\nu \rho} W_{a \mu} +
 i \frac{g \xi}2 (\phi^{\prime \dagger}  \sigma^a \langle \phi \rangle -
    \langle \phi^\dagger \rangle \sigma^a \phi^\prime) \right]^2 \nn \\
            &&      - \frac1{2 \xi} \left[
g_{\mu\nu} \partial^\mu B^\nu + g^{\nu\rho} \Gamma^\mu_{\nu \rho} B_\mu +
 i \frac{ g^\prime \xi}2 (\phi^{\prime \dagger}  \langle \phi \rangle -
    \langle \phi^\dagger \rangle  \phi^\prime) \right]^2
\label{eq:gf}
\eea
that differs from the standard one by the terms\footnote{
The addition of a term proportional to the Christoffel symbol
in the gauge-fixing function of a spin-1 massive particle was considered in 
Ref.\cite{hlz}. However in that work the mechanism of mass-generation for
the spin-1 particle was not addressed.}
proportional to 
$\Gamma^\mu_{\nu \rho}$. In Eq.(\ref{eq:gf}) $\xi$ is the gauge parameter, 
$g,W_a^\mu\:(g^\prime, B^\mu)$ are the coupling constant and  fields
of the $SU(2)\: (U(1))$ group, $\sigma^a$ are the Pauli matrices,
\beq
\Gamma^\mu_{\nu \rho} = \frac12 g^{\mu \alpha} 
       \left( \partial_\rho g_{\alpha \nu} + \partial_\nu g_{\alpha \rho}
         -\partial_\alpha g_{\nu \rho} \right)
\eeq
is the Christoffel symbol, and 
\beq 
\phi^\prime = \left( \begin{array}{c} \phi_+ \\ \frac{ 
\phi_1 + i \phi_2}{\sqrt{2}} \end{array} \right) ~~~~~~~
\langle \phi \rangle = \left( \begin{array}{c} 0 \\ \frac{v}{\sqrt{2}} 
\end{array} \right)
\eeq
with $\phi_1$ the physical Higgs field, $\phi_+$ and $\phi_2$ the unphysical
counterparts of the $W^+$ and $Z$ vector boson and  $v$ the vacuum expectation 
value. 

At the first order in the $\kappa$ expansion, the  interaction
Lagrangian for the tree-level graviton coupling to SM fields, is
\bea
{\cal L}_{int}=-\frac{1}{\Mp}T^{\mu\nu} h_{\mu\nu}
\label{Lint}
\eea
where $T^{\mu\nu}$ is the energy-momentum tensor of matter fields 
obtained from  the Lagrangian in Eq.(\ref{eq:2}) with the gauge-fixing function
given in Eq.(\ref{eq:gf}). 
The mass-scale $\Mp$ appearing in Eq.(\ref{Lint})
is the reduced Plank mass defined as $\Mp \equiv 2/\kappa$.
From Eq.(\ref{Lint}), the  Feynman rules can be derived. 
The ones, in the $R_\xi$ gauge, relevant for our calculation are collected in 
Appendix A. Feynman rules for graviton interactions with vector bosons, in the 
unitary gauge, can be found in Refs.\cite{hlz,np,grw}.

Because of the V-A nature of the charged weak currents, 
the exact expression of the off-diagonal
matrix element (in flavor space) of energy-momentum tensor 
$\hat{T}^{\mu\nu}$ has the following 
structure
\beq
\hat{T}^{\mu\nu} = \frac{iG_F}{16 \pi^2\sqrt{2}} \,
\sum_{i=1}^{12} f_{i}(p,q)\,\bar{u}_2(p_2) O^{\mu\nu}_{i}  u_1(p_1) 
\label{ME}
\eeq
with
\bea
O_1^{\mu\nu}&=&
\left( \gamma^{\mu} p^{\nu}+\gamma^{\nu} p^{\mu}\right) P_L
\nonumber \\
O_2^{\mu\nu}&=&
\left( \gamma^{\mu} q^{\nu}+\gamma^{\nu} q^{\mu}\right) P_L 
\nonumber \\
O_3^{\mu\nu}&=& \eta^{\mu\nu} \, M_+
\nonumber \\
O_4^{\mu\nu}&=& p^{\mu}p^{\nu} \, M_+
\nonumber \\
O_5^{\mu\nu}&=&  q^{\mu}q^{\nu} \, M_+
\nonumber \\
O_6^{\mu\nu}&=&  \left(p^{\mu}q^{\nu}+q^{\mu}p^{\nu}\right)\, M_+
\nonumber \\
O_7^{\mu\nu}&=&  \eta^{\mu\nu} \, M_-
\nonumber \\
O_8^{\mu\nu}&=&  p^{\mu}p^{\nu} \, M_-
\nonumber \\
O_9^{\mu\nu}&=&  q^{\mu}q^{\nu} \, M_-
\nonumber \\
O_{10}^{\mu\nu}&=& \left(p^{\mu}q^{\nu}+q^{\mu}p^{\nu}\right) M_-
\nonumber \\
O_{11}^{\mu\nu}&=&
\frac{m_1 m_2}{m_W^2} 
\left( \gamma^{\mu} p^{\nu}+\gamma^{\nu} p^{\mu}\right) P_R\,
\nonumber \\
O_{12}^{\mu\nu}&=&
\frac{m_1 m_2}{m_W^2} 
\left( \gamma^{\mu} q^{\nu}+\gamma^{\nu} q^{\mu}\right) P_R 
\label{basis}
\eea
where $P_{L,R}=(1\mp \gamma_5)/2$ and $M_{\pm}\equiv m_1P_R\pm m_2P_L$.

The form factors appearing in Eq.(\ref{ME}) are not all 
independent. Indeed, due to the conservation of the energy-momentum tensor
$\partial_{\mu} T^{\mu\nu}(x)=0$ (or equivalently
$\partial_{\nu} T^{\mu\nu}(x)=0$ ), a subset of independent form factors
arise. In momentum space, this relation is translated into 
\bea
q_{\mu} \langle p_2 |~ T^{\mu\nu}(0)~ | p_1\rangle = 0\, ,
\label{WI}
\eea
which reduce the number of form factors $f_i(p,q)$ in Eq.(\ref{ME})
into an independent subset.
Applying Eq.(\ref{WI}) into the right-hand side (r.h.s) 
of Eq.(\ref{ME}), we get the following set of Ward identities (WI)
\bea
(p\cdot q) f_1(p,q)+ q^2 f_2(p,q) &=&0\, ,
\nonumber \\
f_3(p,q)+q^2 f_5(p,q)+(p\cdot q) f_6(p,q)+
\frac{(p\cdot q)}{2 m_W^2} f_{12}(p,q)
&=&0\, ,
\nonumber \\
(p\cdot q) f_4(p,q)+q^2 f_6(p,q)+\frac{(p\cdot q)}{2 m_W^2} f_{11}(p,q)
&=&0\, ,
\nonumber \\
f_2(p,q)+f_7(p,q)+q^2 f_9(p,q) +(p\cdot q) f_{10}(p,q)-\frac{p^2+q^2}{4m_W^2}
f_{12}(p,q)&=&0\, ,
\nonumber \\
f_1(p,q)+(p\cdot q) f_8(p,q)+q^2f_{10}(p,q)
-\frac{p^2+q^2}{4m_W^2} f_{11}(p,q)
&=&0\, ,
\nonumber \\
(p\cdot q) f_{11}(p,q)+q^2 f_{12}(p,q) &=&0\, .
\label{WI2}
\eea
These relations  provide a strong check of our  calculation in  terms of
Feynman diagrams. 

In the case of light external states, the dominant contribution to 
$\hat{T}^{\mu\nu}$ comes from the $O_{1}$ and $O_{2}$ structures.
Indeed the  $O_3$--$O_{10}$ structures have a single chiral suppression 
while  $O_{11}$ and $O_{12}$  are double chirally suppressed.
\begin{figure}[!htb]
\begin{center}
\dofigb{3.1in}{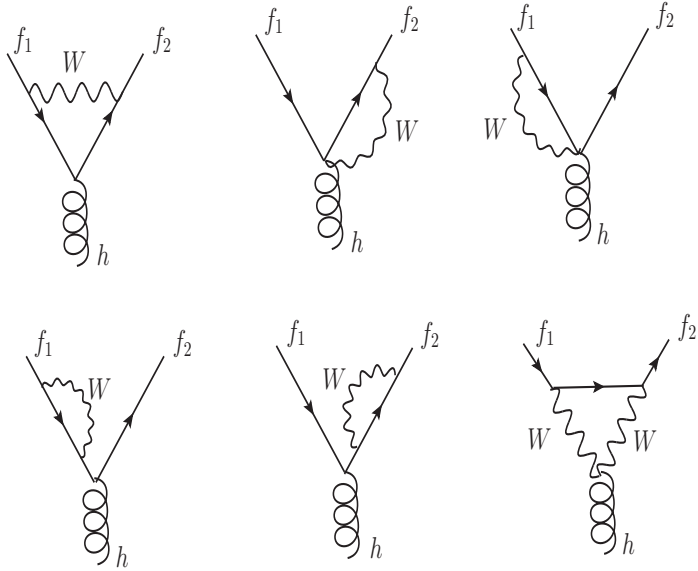}
\end{center}
\caption{\small The 
one-loop Feynman diagrams contributing to the flavor-changing
quark-graviton vertex in the unitary gauge. In the $R_\xi$ gauge there is
also the contribution of diagrams  where the $W$ boson is replaced by its 
unphysical counterpart.}
\label{GAplot}
\end{figure}

The interaction of the graviton with two different fermion fields is described
at the lowest order by the one-loop diagrams drawn\footnote{Feynman diagrams
have been drawn using JaxoDraw \cite{bt}} in Fig.1. To be definite we 
assume the external fermion states to be quarks of different generation, 
then the mixing matrix entering in the $W^+$ and $\phi^+$ vertices is the 
CKM matrix.
We evaluated exactly, namely keeping the complete dependence on the 
$p,q$ momenta and on the internal/external masses, the Feynman
diagrams in Fig.1 to obtain the form factors $f_i(p,q)$ appearing in 
Eq.(\ref{ME}). The calculation has been performed in the 't Hooft-Feynman
gauge ($\xi=1$) and cross-checked in the unitary gauge. While in the
't Hooft-Feynman gauge the result is finite before the application of the GIM
mechanism, in the unitary gauge the GIM is needed to cancel some residual
divergent term. From our result we
explicitly verified all the WI reported in Eq.(\ref{WI2}). It should be pointed
out that to satisfy the WI the $\xi$-dependent term in the $W^+W^-$graviton 
vertex (see Eq.(\ref{xidip}) in Appendix A) was necessary. 

The complete result is too long
to be reported, however;  since we are mainly interested in analyzing the 
effects of the TFCNC  in low energy processes, we report here the leading
term  of the form factors $f_i(p,q)$ in a  $q^2/m_W^2$ and 
$(p\cdot q)/m_W^2$ expansion of the matrix element, 
keeping the dependence on the internal fermion masses exact.
In this approximation, the nonvanishing factors are given by
\bea
f_1(p,q)&=&
- q^2 \sum_f \lambda_f  g_a(x_f)
\nonumber \\
f_2(p,q)&=&
\left(p\cdot q\right)\sum_f \lambda_f g_a(x_f)
\nonumber \\
f_3(p,q)&=&
q^2 \sum_f \lambda_f g_b(x_f)
\nonumber \\
f_5(p,q)&=&
- \sum_f \lambda_f  g_b(x_f)
\nonumber \\
f_7(p,q)&=&
-2 \left( p\cdot q \right) \sum_f \lambda_f g_a(x_f)
\nonumber \\
f_{10}(p,q)&=&
\sum_f \lambda_f g_a(x_f)
\nonumber \\
f_{11}(p,q)&=&
- q^2 \sum_f \lambda_f  g_m(x_f)
\nonumber \\
f_{12}(p,q)&=&
\left(p\cdot q\right)\sum_f \lambda_f g_m(x_f)
\label{FF}
\eea
where $x_f=m_f^2/m_W^2$ with $m_f$ the mass of the fermion running into the 
loop  and $\lambda_f\equiv K_{f1} K^{\star}_{f2}$ (the external quarks are 
assumed to be of the down type),
with $K_{ij}$ the corresponding CKM matrix element. 
The functions $g_i(x)$ appearing in Eq.(\ref{FF}) are given by
\bea
g_a(x)&=& 
\frac{1}{36 \left(x-1\right)^4}
\left[44-194x+243x^2-98x^3 +5x^4
\right.
\nonumber\\
&&~~~~~~~~~~~~~~~+\left. 6x\left(2-15x+10 x^2\right) \log{(x)}\right]
\nonumber \\
g_b(x)&=& 
\frac{1}{6 \left(x-1\right)^4}
\left[8-14x+21x^2-14x^3-x^4
+2x\left(4+3x+2 x^3\right)
\log{(x)}\right]
\nonumber \\
g_{m}(x)&=& 
\frac{1}{72 \left(x-1\right)^6}
\left[6-83x+200x^2+12x^3-142x^4 +7x^5 
\right.
\nonumber\\
&&~~~~~~~~~~~~~~~~ -\left. 12x\left(1+4x-18 x^2-2x^3\right) \log{(x)}\right]
\label{functions}
\eea
It is easy to check that, in this approximation,  
the form factors in Eq.(\ref{FF}) satisfy the WI in Eq.(\ref{WI2}).
From Eqs.(\ref{FF}), (\ref{functions}) one sees that the leading contribution
to $\hat{T}^{\mu\nu}$ is given in terms of two functions, namely 
$g_a(x)$ and $g_b(x)$.
Indeed the explicit expression of the leading terms in $\hat{T}^{\mu\nu}$
is
\bea
\hat{T}^{\mu\nu}&=&\frac{i G_F}{16 \pi^2\sqrt{2}} \,
\sum_f \lambda_f \bar{u}(p_2)
\Big\{
\left(\gamma^{\mu} p^{\nu}+\gamma^{\nu} p^{\mu}\right)P_L
(-q^2)g_a(x_f)+
\left(\gamma^{\mu} q^{\nu}+\gamma^{\nu} q^{\mu}\right)P_L
(p\cdot q)g_a(x_f)
\nonumber\\
&&~~~~~~~~~~~~~~~~~~~~~~~~~~~~~-q^{\mu}q^{\nu} M_+ g_b(x_f)+\eta^{\mu\nu}
     \Big[-2(p\cdot q)M_-g_a(x_f) +q^2M_+g_b(x_f)\Big]
\nonumber\\
&&~~~~~~~~~~~~~~~~~~~~~~~~~~~~~+\left(p^{\mu}q^{\nu}+p^{\nu}q^{\mu}\right)M_
 -g_a(x_f)\Big\}u(p_1)\, ,
\label{Tleading}
\eea
and gauge invariance, $q_{\mu}T^{\mu\nu}=0$, is easily verified using 
$q_{\mu} \bar{u}(p_2) \gamma^{\mu}P_L u(p_1)=\bar{u}(p_2) M_-u(p_1)$.

In Appendix B we present  $\hat{T}^{\mu\nu}$ in the approximation
$(p\cdot q)= p^2 +q^2 =0$ keeping the full dependence on $q^2$ and the
internal masses.

\section{Flavor-changing gravitational couplings}
In this section we analyze the matrix element of the TFCNC coupled to 
an external gravitational source  at the tree-level.
We consider the following gravitational scattering 
\bea
f_1(p_1) + T^{\rm ext} \to f_2(p_2) + T^{\rm ext}
\eea
where $T^{\rm ext}_{\mu\nu}$ indicates the energy-momentum tensor of
the external source, assumed to be conserved 
($\partial^{\mu} T^{\rm ext}_{\mu\nu}=0$),
and $f_{1,2}(p_{1,2})$ are two fermions of different flavor and masses, 
with associated four-momenta $p_{1,2}$ respectively.
In momentum space, the corresponding one-graviton exchange amplitude is
given by
\bea
{\cal M}=\frac{1}{\Mp^2}\, \hat{T}_{\mu\nu}~ 
P_h^{\mu\nu\alpha\beta}(q^2) ~ \hat{T}^{\rm ext}_{\alpha\beta}
\label{Mamp}
\eea
where $P_h^{\mu\nu\alpha\beta}(q^2)$ is the graviton propagator in momentum
space, with $q=p_1-p_2$, 
and $\hat{T}^{\rm ext}_{\mu\nu}$ is the Fourier transform of 
$T^{\rm ext}_{\mu\nu}(x)$.
It is understood that all indices are 
contracted with the Minkowski metric $\eta_{\mu\nu}$.
In the Einstein theory, 
the graviton propagator, in a covariant gauge, is given by
\bea
P_h^{\mu\nu\alpha\beta}(q^2)=\frac{i}{q^2-i\varepsilon}\frac{1}{2}\left(
\eta^{\mu\alpha}\eta^{\nu\beta} + 
\eta^{\nu\alpha}\eta^{\mu\beta} - \eta^{\mu\nu}\eta^{\alpha\beta}
+ Q^{\mu\nu\alpha\beta}(q)\right)
\label{EinstProp}
\eea
where the last term  $Q^{\mu\nu\alpha\beta}(q)$, which is gauge dependent, 
is a tensor made by linear combinations
of an even number of $q$ momenta with open indices, 
such as for instance $q^{\mu}q^{\nu} \eta^{\alpha\beta}$ or
$q^{\mu}q^{\nu} q^{\alpha}q^{\beta}$. It is also possible to set 
$Q^{\mu\nu\alpha\beta}=0$ by a particular gauge choice \cite{vDV}.
Nevertheless, due to  the conservation of the
energy-momentum tensor, the 
contribution of $Q^{\mu\nu\alpha\beta}(q)$ vanishes when contracted with 
$\hat{T}_{\mu\nu}$ or $\hat{T}^{\rm ext}_{\alpha\beta}$, leading
to a gauge-invariant result.
By using the energy-momentum conservation, we can write
\bea
{\cal M}=\frac{i}{\left(q^2-i\varepsilon\right)2 \Mp^2}
\left[2
\hat{T}^{\mu\nu}\, \hat{T}^{\rm ext}_{\mu\nu}-
C\, \hat{T}^{\mu}_{\mu} \hat{T}^{\rm ext~ \nu}_{\nu}\right]\, .
\label{Mc}
\eea
where we introduce, for later reference, the numerical factor $C$ that
in the present case (massless graviton) is equal to one.
Because of the fact that the r.h.s. of 
Eq.(\ref{Mc}) is not vanishing, the scattering of a fermion on 
an external gravitational field can induce a FC transition.

Despite the presence of the $1/q^2$ pole in Eq.(\ref{Mc}), 
the FC gravitational transition turns out to be local.
Indeed
by using the leading contributions from Eq.(\ref{Tleading}) in Eq.(\ref{Mc})
we find
\bea
{\cal M} &=& \frac{ G_F  }{16\pi^2 \sqrt{2}\Mp^2} \sum_f \lambda_f
\bar{u}(p_2)\left\{ \Big(\gamma^{\mu} p^{\nu}+\gamma^{\nu} p^{\mu}\Big)P_L 
g_a(x_f) - \eta^{\mu\nu}M_{+}\Big(C\,g_a(x_f)-(\frac{3}2\,C -1 ) g_b(x_f)\Big)
\right. \nonumber \\
&& \left. ~~~~~~~~~~~~~~~~~~~~~~~~~~~~~~~~~~
+ \eta^{\mu\nu} M_{-} \frac{(p\cdot q)}{(q^2-i\varepsilon)} g_a(x_f) 
\left[2+C\left(-2\right)\right]
\right\} u(p_1)\, 
\hat{T}^{\rm ext}_{\mu\nu}
\label{Mgeneral}~.
\eea
The second line in Eq.(\ref{Mgeneral}) is zero if C=1 (massless graviton) 
showing that in the Einstein theory of 
general relativity  the $1/q^2$ pole  cancels out.
This is a general result that holds also in the exact case, as can
be easily proved  using the WI in Eq.(\ref{WI2}). 

As we can see from Eq.(\ref{Mgeneral}), gravity and weak interactions induce 
at one loop an effective local interaction for the $\Delta F=1$ flavor 
transitions.
The leading contribution to the corresponding effective Hamiltonian 
is given by  local operators of dimension eight.
The characteristic scale of this effective theory  is
$\Lambda=\sqrt{M_{P} m_W}\sim 10^{10}$ GeV, which is 6 orders of magnitude
below the grand unified theory (GUT) scale.
However, with respect to the known dimension six operators in the 
$\Delta F=1$ effective Hamiltonian of weak interaction the TFCNC gravitational
operators induce negligible effects because the scale $\Lambda$ is much higher
and moreover it enters in Eq.(\ref{Mgeneral}) at the fourth power.

The fact that the $1/q^2$ pole cancels out in 
the amplitude ${\cal M}$ in Eq.(\ref{Mc})
for a massless graviton, can be simply understood in terms of 
angular-momentum conservation.
Let us suppose that the numerator of ${\cal M}$ does not vanish 
in the $q^2\to 0$ limit. If this would be the case, then TFCNC could 
be directly coupled to a on-shell massless graviton ($q^2=0$).
In particular, a nonvanishing matrix element 
for the fermion decay $f_1\to f_2 G$ would be possible,
where $G$ is an on-shell graviton and $m_{f_1} > m_{f_2}$. 
However, the decay
$f_1\to f_2 G$ is forbidden by angular-momentum
conservation as can be easily understood looking, in the rest frame of the
decaying particle,  at the conservation of the angular momentum along
the momentum direction of the two final states.
In this frame one can see that angular momentum conservation 
is unbalanced along this direction, since a massless graviton
carries only helicity states $\pm 2$, while fermions can only have 
helicity states $\pm 1/2$. Therefore, in the case of massless graviton,
only contact terms in the amplitude ${\cal M}$ are allowed.
Instead the fermion decay
$f_1\to f_2 G$ does not vanish  if the graviton has a small mass. 
Indeed, the spin content of a massive graviton contains
five independent polarization states, including, among the spin-2 and
spin-1 polarizations, also a spin-0 one. 
Then, the on-shell transition $f_1\to f_2 G$ is allowed by 
angular-momentum 
conservation due to the presence of the spin-0 graviton polarization.

We consider now a scenario in which the graviton has a very small
mass $m_G$.
In this case, the corresponding graviton propagator in the unitary gauge
is given by \cite{vDV}
\bea
P_{G}^{\mu\nu\alpha\beta}(q^2)= \frac{i}{q^2-m_G^2-i\varepsilon}
\frac{1}{2}\left(
\tilde{\eta}^{\mu\alpha}\tilde{\eta}^{\nu\beta} + 
\tilde{\eta}^{\nu\alpha}\tilde{\eta}^{\mu\beta} - 
\frac{2}{3}\tilde{\eta}^{\mu\nu}\tilde{\eta}^{\alpha\beta}
\right)
\label{PropGm}
\eea
where $\tilde{\eta}^{\mu\nu} \equiv \eta^{\mu\nu} - q^\mu q^\nu/m_G^2$.
The terms proportional to $q/m_G$, that in principle can be very enhanced, 
actually  vanish when  contracted with $T_{\mu\nu}$ and therefore do not
contribute to the analogous transition amplitude ${\cal M}$ in Eq.(\ref{Mc}).
We assume here that $m_G$ is small enough 
that the corresponding Newton potential ($V(r)\sim \frac{1}{r}e^{-r m_G}$)
is not affected by its effect. This is clearly true for distances 
$r\ll 1/m_G$, where $V(r)\sim 1/r$.

We recall that, apart from terms proportional to the momentum $q$, the 
massive-graviton propagator in Eq.(\ref{PropGm}) differs from the one
of the Einstein theory in Eq.(\ref{EinstProp}) by a finite term
proportional to $\eta^{\mu\nu}\eta^{\alpha\beta}$ which does not vanish in the
limit $m_G\to 0$.
This is a consequence of the fact that the spin-0 component of the 
graviton field does not decouple in the limit $m_G \to 0$, giving
rise to a real graviton-mass discontinuity.
This phenomenon was discovered by van Dam, Veltman 
\cite{vDV} and Zakharov \cite{zak} (vDVZ), 
by analyzing the one-graviton exchange amplitude.
In particular, they found that if the graviton has a small mass, 
no matter how small, a finite difference in the 
deflection angle for the light-bending from the sun would be predicted
with respect to the massless case. Then, since this prediction
is out of 25\% from the measured value (which is in agreement with the general
relativity predictions), they concluded that the massive gravity theory
can be ruled out by solar system observations. This conclusion relies
on the fact that the terms singular in $m_G$  vanish in the one-graviton 
approximation.  However, there are criticisms on the validity of this 
approximation based on the observation that higher order corrections can be
singular in the graviton mass and therefore cannot be neglected  
\cite{vainsh,ddgv}.

Here we show that there exists another aspect of the vDVZ discontinuity, 
to our knowledge not considered so far, concerning the
flavor-changing contribution to the Newton potential in the massive gravity
theory. In order to see that, let us consider
the amplitude in Eq.(\ref{Mamp}) in the 
case of one massive-graviton exchange. It
can  be easily obtained from Eq.(\ref{Mgeneral}) with $C=2/3$ giving
\bea
{\cal M} &=& \frac{ G_F  }{16\pi^2 \sqrt{2}\Mp^2} \sum_f \lambda_f
\bar{u}(p_2)\left\{ \frac{q^2}{(q^2-m_G^2-i\varepsilon)}
\left( \Big(\gamma^{\mu} p^{\nu}+\gamma^{\nu} p^{\mu}\Big)P_L 
 - \frac23 \eta^{\mu\nu}M_{+}\right) g_a(x_f)
\right. \nonumber \\
&& \left. ~~~~~~~~~~~~~~~~~~~~~~~~~~~~~~~~~~
+ \frac23 \eta^{\mu\nu} M_{-} \frac{(p\cdot q)}{(q^2-m_G^2-i\varepsilon)} 
g_a(x_f) \right\} u(p_1)\, \hat{T}^{\rm ext}_{\mu\nu}
\label{Mgmass}~.
\eea
A remarkable aspect of Eq.(\ref{Mgmass}) is that the
$1/q^2$ pole in the last term does not vanish
in the limit $m_G\to 0$. As a consequence 
it generates an off-diagonal contribution (in flavor space) to the Newton
potential. In order to see that, one has
to first look at the nonrelativistic limit of Eq.(\ref{Mgmass}) in the case
of an external heavy gravitational source. This should be compared
with the Born approximation to the corresponding scattering amplitude
in nonrelativistic quantum mechanics written in terms of the potential
function $V(r)$. 
Then, the matrix element of the Newton potential $V(r)$
(generated by a heavy massive particle $M$) between initial and final
fermions states is
\beq
\langle j\, |\, V(r)\, |\, i\, \rangle
=-\frac{G_N M}{r} e^{-r\, m_G}\left(m_i\, \delta_{ij}
+\left(m_i-m_j\right)\, \Delta_{ij}\right) 
\label{Newton}
\eeq
with
\beq
\Delta_{ij}=\frac{G_F (m_i^2-m_j^2)}{32 \pi^2 \sqrt{2}}\sum_f
K_{fi} K^{\star}_{f j} \, g_a(x_f)\, ,
\label{Newt}
\eeq
where $G_N$ is the Newton constant for the massive-graviton case \cite{vDV}
and $\delta_{ij}$ is the delta-function in flavor space,
the indices $i$ and $j$ standing for the ingoing and outgoing
fermion states respectively.
The discontinuity in $m_G$ is manifest in the fact that
in the limit $m_G\to 0$ the $\Delta_{ij}$ term does not vanish,
while the $\Delta_{ij}$ is strictly zero in the case of massless graviton.
Notice that the FC contribution to the Newton potential vanishes in the case
of equal masses ($m_i=m_j$) and its attractive or
repulsive nature is related to the sign of the $\sum_f
K_{fi} K^{\star}_{f j} \, g_a(x_f)$ quantity.
 It should be recalled that the result of 
Eq.(\ref{Newton}) has been derived in the one-graviton approximation
and therefore all the criticisms with respect to the vDVZ discontinuity
apply also to it \cite{vainsh,ddgv}.
\section{Flavor-changing processes with massive spin-2 particles}
In this section  we analyze a few 
applications of the results derived in Secs. 2 and 3
related to processes that are characterized by the kinematical 
regimes of small and large values of $|q^2|/m_W^2$.
At this aim, we consider a scenario containing an elementary massive 
spin-2 particle in the spectrum
coupled to a conserved energy-momentum tensor  $T_{\mu\nu}(x)$
via the interaction Lagrangian
\bea
{\cal L}_{\rm eff}=-\frac{1}{\Lambda}\,T_{\mu\nu}(x) G^{\mu\nu}(x)\, ,
\label{Lspin2}
\eea
where $G_{\mu\nu}(x)$ represents the massive spin-2 field
and  $\Lambda$ is a mass-scale free parameter having no relation with 
the Plank mass $\bar{M}_P$.
For the spin-2 free Lagrangian we take the Pauli-Fierz action which is
the ghost-free linearized action for a massive spin-2 particle \cite{vDV}.
Since we will consider here only the first order in the 
$1/\Lambda$ expansion, self-interactions of the spin-2 fields 
can be neglected and the $T_{\mu\nu}(x)$ is reduced 
to the energy-momentum tensor of matter fields in flat space-time.
Then, the off-diagonal one-loop correction to the Lagrangian 
in Eq.(\ref{Lspin2}) can be easily generalized
from the gravitational case  by replacing $\bar{M}_P \to \Lambda$.

We consider first the case of a decay process
\bea
f_1(p_1)\to f_2(p_2)\,  G(q)
\label{fdecay}
\eea
that can  be considered as the spin-2 counterpart 
of the well-known fermion radiative decay $f_1\to f_2 \gamma$.
In Eq.(\ref{fdecay}) $f_1$ and $f_2$ stand for 
two generic different fermions of the same charge  with $p_{1,2}$
the corresponding momenta, $q =p_1-p_2$ and all the external particles
are assumed to be much lighter than the W boson and the fermions running
into the loop. 
The corresponding decay amplitude can be written as:
\bea
M(f_1\to f_2 G)=-\frac{ i m_G^2 G_F}{16 \pi^2\sqrt{2}\Lambda} \,
\sum_f \lambda_f g_a(x_f)\, \bar{u}_2(p_2)
\Big[\gamma^{\mu}p^{\nu}+\gamma^{\nu} p^{\mu} \Big] P_Lu_1(p_1)\, 
\epsilon^i_{\mu\nu}(q)
\label{Mbsg}
\eea
where $\epsilon^i_{\mu\nu}(q)$ is the polarization tensor of the 
spin-2 particle of mass $m_G$ with the index $i=1,5$  labeling the 
five independent polarizations and
the function $g_a(x)$ is  defined in Eq.(\ref{functions}). 
In obtaining Eq.(\ref{Mbsg}) 
we have used the fact that for an on-shell spin-2 particle one has
$q_{\mu} \epsilon^{\mu\nu}(q)=0$ and $\eta_{\mu\nu} \epsilon^{\mu\nu}(q)=0$.

We point out that in the limit $m_G\to 0$ the amplitude in Eq.(\ref{Mbsg})
seems to vanish, being
proportional to $m_G^2$. However, the sum over
polarizations of a massive  spin-2 particle \cite{vDV,grw}
\bea
\sum_{i=1}^5 
\epsilon^i_{\mu\nu}(q) \epsilon^{i~ \dag}_{\alpha\beta}(q)
&=&
\frac{1}{2}\left(
\tilde{\eta}^{\mu\alpha}\tilde{\eta}^{\nu\beta} + 
\tilde{\eta}^{\nu\alpha}\tilde{\eta}^{\mu\beta} - 
\frac{2}{3}\tilde{\eta}^{\mu\nu}\tilde{\eta}^{\alpha\beta}
\right)
\label{projector}
\eea
contains terms singular as $m_G\to 0$. Thus, after summing over all the 
polarizations,  averaging over the initial ones, and integrating over the 
phase space, the decay width, assuming $f_2$ massless, reads: 
\bea
\Gamma(f_1\to f_2 G)=\frac{G_F^2 m_1^7  f(x_G)}{192\, (2\pi)^5\, \Lambda^2}
\,\Big|\sum_f \lambda_f g_a(x_f)\Big|^2\,
\label{GammafG}
\eea
where $x_G=m_G^2/m_1^2$, and $f(x)=(1-x)(1-
\frac{3}{2}(x+x^2+x^4)+\frac{7}{2}x^3)$
absorbs the matrix element and phase space corrections.

These results can have an application in the framework 
of quantum gravity propagating in large extra dimensions \cite{ADD,AADD}.
In \cite{ADD,AADD}, it was pointed out that if compactified extra dimensions 
exist, with only gravity propagating in the bulk, the fundamental scale
of quantum gravity could be much lower than the Plank scale $M_P$.
In this scenario, the standard Newton constant $G_N$ in (3+1)-dimensional
space is related to the corresponding Plank scale $M_D$ in 
($D=4+\delta$)-dimensional space, by
\bea
G_N^{-1}=8\pi R^{\delta}\, M_D^{2+\delta}
\eea
where $R$ is the radius of the compactified manifold assumed here to be on 
a torus. If one requires $M_D \sim$ TeV, present tests on gravity law 
imply that $\delta \ge 2$.

After integrating out compact extra dimensions, the effective low energy 
theory describes an almost continuous spectrum of massive spin-2 particles, 
which are excitations of the standard graviton field. Then, each massive 
spin-2 field is coupled to the matter field by Eq.(\ref{Lspin2}), where
the energy scale $\Lambda$ corresponds to the reduced Plank mass $\bar{M}_P$.
In the case of $M_D\sim$ TeV and $\delta< 4$, the mass splitting between the 
Kaluza-Klein (KK) 
excitations is of the order of KeV, and the KK spectrum can be 
approximated as a continuous. In this case the number density of modes
($dN$) between $m_G$ and $m_G+dm_G$ of KK spin-2 masses is \cite{grw}
\bea
dN=S_{\delta-1}\frac{\bar{M}_P^2}{M_D^{2+\delta}}\, m_G^{\delta-1}\, dm_G,
\eea
where $S_{\delta-1}$ is the surface of a unit-radius sphere in 
$\delta$ dimensions which is given by $S_{\delta-1}=2\pi^n/(n-1)!$
and $S_{\delta-1}=2\pi^n/\prod_{k=0}^{n-1}(k+\frac{1}{2})!$ for
$\delta=2n$ and $\delta=2n+1$, with $n$ integer, respectively.

In this framework, we consider the {\it inclusive} flavor-changing
graviton decay $f_1\to f_2 G$, where
$G$ stands for any KK massive spin-2 gravitons with mass $m_G < m_1$.
In this case, the corresponding decay width
is obtained by multiplying 
Eq.(\ref{GammafG}) for $dN$, with $\Lambda$ replaced by $\bar{M}_P$,
and integrating it over all the allowed kinematical 
range of $m_G$.
In particular, for the inclusive decay we get
\bea
\sum_G \Gamma(f_1\to f_2 G)=\frac{G_F^2 m_1^{7+\delta}  
S_{\delta-1}\, I(\delta)}{192\, (2\pi)^5\, M_D^{2+\delta}}
\,\Big|\sum_f \lambda_f g_a(x_f)\Big|^2\,
\label{GammafGED}
\eea
where $I(\delta)=960/
(\delta(2+\delta)(6+\delta)(8+\delta)(10+\delta))$ and $\delta \ge 2$.
As we can see from  Eq.(\ref{GammafGED}), the integration
over the number of KK states cancels the $1/\bar{M}_P^2$ suppression 
factor of the single graviton emission.
We stress that the final KK gravitons are detected as missing energy,
since for laboratory experiments they can be approximated as stable particles
\cite{AADD}.
This is due to the fact that the decay width of a single KK graviton 
is strongly suppressed by $1/\bar{M}_P^2$.

For a numerical evaluation of Eq.(\ref{GammafGED}) we restrict
ourselves to the inclusive B-meson decay $B_d\to X_s G$, where the standard 
GIM suppression is enhanced due to the contribution of the 
top-quark running in the loop. In particular, for $\delta=2$,
the corresponding branching-ratio, normalized to the experimental 
$BR(B\to X \bar{\nu} e) \simeq 10.4\%$ of the semileptonic decay,
is 
\bea
\sum_G 
BR(B_d\to X_s G)\simeq 10^{-13}\, \left(\frac{{\rm TeV}}{M_D}\right)^4
\left(\frac{m_b}{4.3\, {\rm GeV}}\right)^4,
\eea
where $X_s$ stands for any hadronic state 
containing an $s$-quark. The above result corresponds to a top-quark mass 
$m_t=171.2$ GeV.

Next we consider the decay of a massive spin-2 particle
$G\to f_2 \bar{f_1}$, where $f_{1,2}$ are two fermions
of different flavor.  In the approximation 
of neglecting the final fermion masses the
corresponding amplitude can be obtained from
Eq.(\ref{Tleading}) by setting in it $(p\cdot q) =M_+=M_- =0$ 
and replacing the function 
$g_{a}(x)$ with the function $G_{a}(x,y)$ provided in Appendix B,
where the full $q^2$ dependence in the form factors
is retained, or
\bea
\!\!\!\!\!\!\!\!\!\!
M(G\to f_2 \bar{f_1})\!\!
&=&\!\!-\frac{ i\, G_F m_G^2}{16 \pi^2\sqrt{2}\, \Lambda}
\, \sum_f \lambda_f\, G_a(x_f,x_G)
\bar{u}_2(p_2)
\Big[\gamma^{\mu} p^{\nu}+\gamma^{\nu} p^{\mu}\Big] P_L\, v_1(p_1)\, 
\epsilon^i_{\mu\nu}(q)
\, .
\label{MGdecay}
\eea
In Eq.(\ref{MGdecay}) $x_G=m_G^2/m_W^2$, $p=p_2-p_1$ and the $v_1(p_1)$ is the 
spinor of the antifermion associated with $\bar{f}_1$.
Then, for the corresponding decay width we obtain
\bea
\Gamma(G\to f_2 \bar{f}_1)=\Gamma_f\,
\frac{G_F^2 m_G^4}{64\pi^4}
\, \Big|\sum_f \lambda_f\, G_a(x_f,x_G)\Big|^2
\label{GammaGdecay}
\eea
where 
\bea
\Gamma_f=\frac{m_G^3 N_c}{160\pi\, \Lambda^2}
\eea
is the tree-level decay width of a massive spin-2 particle 
into a pair of massless fermions of the same flavor, with $N_c$ the color 
factor.

Because of the presence of the $G_F^2 m_G^4$ term in the numerator of 
Eq.(\ref{GammaGdecay}), one would na\"ively expect that when  
$m_G\gg m_W$, the one-loop  decay would be strongly enhanced.
However, this is not the case. Indeed,
by using the asymptotic expansion of the 
 ${\rm B_0}(y)$ and ${\rm C_0}({x,y})$ functions
at large values of $y$ \cite{rd}, we obtain
\bea
{\it G_a}(x,y)&=&
\frac{1}{y}\left[{\it F}(x)-\frac{x}{2}
\Big(\log (y)-i\pi\Big)\right] + \bar{G}(y)
+{\cal O}(\frac{1}{y^2})~,
\label{gga} \\ 
 {\it F}(x)&=& 
\left(\frac{3x\left(3-x\right)}{4(1-x)}+
\frac{x\left(8-6\, x+x^2\right)\log (x)}{2\left(x-1\right)^2}\right)
-4\Big({\rm Li}_2(1-x)- \frac{\pi^2}6\Big)
\eea 
where  ${\rm Li}(\rm x)$ stands for the usual dilogarithm function
and $\bar{\rm G}(y)$ is a pure function of $y$
whose leading term is proportional to $\log^2 (y)/y$, showing that
no powerlike enhancement  is present in the 
FC decay when $m_G\gg m_W$. Furthermore, 
due to the unitarity of CKM, the contribution of $\bar{\rm G}(y)$ vanishes 
when the sum over all internal fermion masses is performed, so that
in the limit $m_G\gg m_W$ the decay width reads:
\bea
\left. \Gamma(G\to f_2 \bar{f}_1) \right|_{m_G \gg m_W} =\Gamma_f\,
\frac{G_F^2 m_W^4}{64\pi^4} \Big|\sum_f \lambda_f 
\left(F(x_f)-\frac{x_f}{2}
\left(\log(\frac{m_G^2}{m_W^2})-i \pi\right)\right)
\Big|^2\, .
\label{BR}
\eea

As a 
last example, we consider the high-energy limit of the following FC fermion
scattering process
\bea
f_i(p_1) X(q_1) \to f_j(p_2) X(q_2)
\label{FCscattering}
\eea
induced by the one-graviton exchange amplitude, 
where $f_{i,j}$ stands for two fermions of 
different flavor and $X$ indicates a generic particle.  
Because of the fact that gravity does not change flavor at 
tree level, only the t-channel diagram will contribute 
to the above scattering (where $t\equiv (p_1-p_2)^2$ ), provided
the $X$ particle is different from the initial and final fermions.
In the fermion massless limit, only the
first term in Eq.(\ref{Mgeneral}), proportional to the $G_a(x,y)$ 
function, will contribute to the matrix element and therefore,
regardless the graviton is assumed massive or 
massless, in the region of large $t$ values  ($|t|\gg m_W^2$),  
the following relation among cross sections holds:
\bea
\frac{d \sigma}{dt}^{f_i \to f_j }
&\simeq&
\frac{d \sigma}{dt}^{f_i \to f_i } \times
\frac{G_F^2 m_W^4}{64\pi^4} \Big|\sum_f \lambda_f 
\Big[F(x_f)-\frac{x_f}{2} \log(\frac{-t}{m_W^2})\Big]
\Big|^2\, ,
\label{sigmarel}
\eea
where
$\frac{d \sigma}{dt}^{f_i\to f_i}$ stands for the diagonal part (in flavor)
of the tree-level differential cross section of 
the process $f_i X \to f_i X$ mediated by the one-graviton exchange diagram.

\section{Conclusions}
We computed the one-loop flavor-changing quark-graviton vertex verifying
all the WI induced by the conservation of energy-momentum tensor. 
The calculation was performed in the $R_\xi$ gauge with $\xi =1$,
using a modified version of the 't-Hooft gauge-fixing function,  and 
cross-checked in the unitary gauge.
We found that the corresponding form factors turn out to be strongly 
GIM suppressed when the internal fermion masses are much smaller than the 
W mass. These results can be easily 
generalized to the corresponding FC lepton-graviton vertices, provided
the neutrinos acquire mass of the Dirac type.

We investigated  the case of a flavor-changing fermion-graviton 
vertex coupled to an external gravitational source. 
We show that, due to the angular-momentum conservation, gravity and
weak interactions induce at one-loop an effective 
local interaction for the $\Delta F=1$ transitions.
At low energy, the corresponding effective Hamiltonian is 
given by local operators of dimension eight suppressed by a characteristic 
scale of order $\Lambda =\sqrt{\bar{M}_P m_W}\sim 10^{10}$ GeV.
Thus  gravitational FC effects  are completely negligible when 
compared to the known  contribution of $\Delta F=1$ dimension six operators 
of the SM.

We showed that, in the case of massless graviton,  the locality of the FC 
gravitational interaction is related to 
the cancellation of $1/q^2$ pole in  the corresponding one-graviton exchange 
amplitude. The latter  does not take place
if the graviton has a small mass signaling another aspect of
the known vDVZ discontinuity in the graviton mass. Indeed we showed that
in the graviton-massive case  the Newton potential
acquires a FC contribution that vanishes in the limit of equal external masses.

As a few  applications of our results
we analyzed a new physics scenario containing
massive spin-2 particles coupled to the SM fields.
In this framework, we calculated the width of the FC
decay $f_1\to f_2 G$  and 
the decay of a heavy spin-2 particle in two fermions of different 
flavor, $G\to f_2 \bar{f}_1$. 
We also considered
the scenario of quantum gravity propagating in large extra dimensions.
In this framework, we evaluated the inclusive decay width for the process
$f_1\to f_2 G$, where $G$ stands for any Kaluza-Klein spin-2 graviton, 
and estimated the branching ratio for the inclusive $B_d$-meson decay 
$B_d\to X_s G$.
Finally, as a consistency test of our results, 
we studied the asymptotic behavior for large 
spin-2 masses of the decay $G\to f_2 \bar{f}_1$ and the high-energy limit of 
the gravitational scattering
$f_1 X \to f_2 X$ (with $X\neq f_{1,2}$) with respect to the GIM mechanism.
We explicitly check that also in the asymptotic limit, 
the GIM mechanism acts in the usual way strongly suppressing the process
for small internal fermion masses.

The results presented for the massive graviton can be easily generalized to 
the case of a graviscalar, $\phi_S$, whose interaction with 
the SM fields can be described by
\bea
{\cal L}_{\rm int}=-\frac{1}{\Lambda_S}\,T_{\mu\nu}(x) \phi_S(x) \eta^{\mu\nu}~.
\nn
\eea
In particular, the counterpart of Eq.(\ref{Newton}) for the graviscalar 
reads:
\bea
\langle j\, |\, V(r)\, |\, i\, \rangle
=-\frac1{\Lambda_S^2}\frac{M}{4 \pi r} e^{-r\, m_{\phi_S}}\left(m_i\, \delta_{ij}
+\left(m_i-m_j\right)\, \Delta^S_{ij}\right)  \nn
\eea
with $\Delta^S_{ij} = 6 \Delta_{ij}$. The width of the FC decay 
$f_1 \to f_2 \phi_S$ neglecting the mass of the produced particles
is instead:
\bea
\Gamma(f_1\to f_2 \phi_S)=\frac{G_F^2 m_1^7 }{32\, (2\pi)^5\, \Lambda_S^2}
\,\Big|\sum_f \lambda_f g_a(x_f)\Big|^2 ~.
\nn
\eea
Finally, because the coupling between a graviscalar and the fermions is
proportional to the fermion  mass, the decay width of a graviscalar into
two massless fermions is null.
\section*{Acknowledgments}
We thank G.~Dvali, G.~Giudice,  R.~Rattazzi and G.~Veneziano for useful 
discussions. This work was supported in part by an EU Marie-Curie Research 
Training Network under Contract No. MRTN-CT-2006-035505 (HEPTOOLS) and by 
the INTAS Project No. 05-1000008-8328.

\section*{Appendix A}
\begin{figure}[!htb]
\begin{center}
\dofiga{10cm}{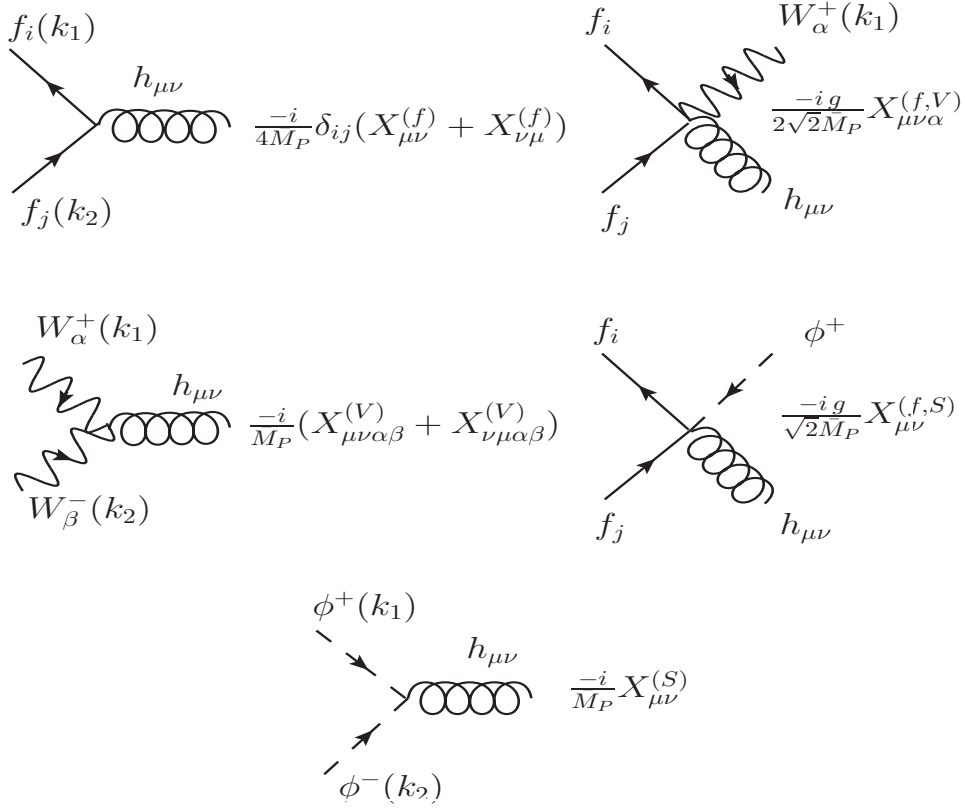}
\end{center}
\caption{\small Three and four-point vertex Feynman rules relevant for our
calculation. The particle momenta are assumed to flow along the
direction of the arrows. For the meaning of the symbols see text.}
\label{FeynRul}
\end{figure}

In this appendix we report the Feynman rules 
for gravitational interactions with SM fields that are 
relevant for the processes considered in this article. 
They are presented in Fig.\ref{FeynRul}.  

\noindent The symbols used in 
Fig.\ref{FeynRul} are defined as:
\bea
X^{(f)}_{\mu\nu}&=&
\gamma_{\mu}\left(k_{1\nu}+k_{2\nu}\right)-
\eta_{\mu\nu}\left(\ksl_1 +\ksl_2-2 m_{i}
\right)
\\
X^{(V)}_{\mu\nu\alpha\beta}&=& 
\frac{1}{2}\eta_{\mu\nu} k_{2\alpha}k_{1\beta} +
\eta_{\alpha\beta} k_{1\mu}k_{2\nu}-\eta_{\mu\beta} k_{1\nu}k_{2\alpha}
-\eta_{\mu\alpha} k_{2\nu}k_{1\beta}   \nonumber \\
&&+ \left( k_1\cdot k_2 +m_W^2 \right)\left(
\eta_{\mu\alpha}\eta_{\nu\beta}
-\frac{1}{2}\eta_{\mu\nu}\eta_{\alpha\beta}\right)  \nn \\
&& + \frac1{\xi} \left[\frac{1}{2}\eta_{\mu\nu} \left( k_{1\alpha}k_{1\beta} +
k_{2\alpha}k_{2\beta} +k_{1\alpha}k_{2\beta} \right) - 
\eta_{\nu\beta} k_{1\alpha}k_{1\mu} - \eta_{\nu\alpha} k_{2\beta}k_{2\mu} 
\right]
\label{xidip} 
\\
X^{(S)}_{\mu\nu}&=& \left( k_1\cdot k_2 +\xi m_W^2 \right) \eta_{\mu\nu} - 
k_{1\nu}k_{2\mu} - k_{1\mu}k_{2\nu}
\\
X^{(f,V)}_{\mu\nu\alpha}&=& K_{ij} \left(
2\gamma_{\alpha}\eta_{\mu\nu}
-\gamma_{\mu}\eta_{\nu\alpha}- \gamma_{\nu}\eta_{\mu\alpha} \right)   P_L
\\
X^{(f,S)}_{\mu\nu}&=& 
K_{ij} \left( \frac{m_j}{m_W} P_R - \frac{m_i}{m_W} P_L \right) 
\eta_{\mu\nu}~.
\eea

\section*{Appendix B}
In this appendix we discuss the form taken by $\hat{T}^{\mu\nu}$
in the limit $(p\cdot q) =p^2 +q^2 =0$, i.e. when terms proportional
to $m_{1,2}^2$ are neglected. In this limit, from the WI on 
Eq.(\ref{WI2}),  only five form factors survive
and $\hat{T}^{\mu\nu}$ can be cast in the following form
\bea
\hat{T}^{\mu\nu}=\frac{i G_F}{16 \pi^2\sqrt{2}}\, 
\sum_{i=1}^5 F_i(p,q) \bar{u}_2(p_2) \widehat{O}^{\mu\nu}_{i} u_1(p_1)
\label{matrixel}
\eea
where
\bea
\widehat{O}_1^{\mu\nu}&=&
  \left(\gamma^{\mu} p^{\nu}+\gamma^{\nu} p^{\mu} \right) P_L  - 
\frac{p^{\mu}q^{\nu}+q^{\mu}p^{\nu}}{q^2} M_-\, ,
\nonumber \\
\widehat{O}_2^{\mu\nu}&=& \left(-\eta^{\mu\nu}q^2+q^{\mu}q^{\nu}\right) M_+\, ,
\nonumber \\
\widehat{O}_3^{\mu\nu}&=& \left(-\eta^{\mu\nu}q^2+q^{\mu}q^{\nu}\right) M_-\, ,
\nonumber \\
\widehat{O}_4^{\mu\nu}&=& \left(p^{\mu}p^{\nu}\right) M_+\, ,
\nonumber \\
\widehat{O}_5^{\mu\nu}&=& \left(p^{\mu}p^{\nu}\right) M_-.
\eea
with $F_1(p,q)=f_{1}(p,q), \: F_2(p,q)=f_{5}(p,q), \:
F_3(p,q)=f_{9}(p,q),\: F_4(p,q)=f_{4}(p,q)$, and 
$F_5(p,q)=f_{8}(p,q)$.
In this basis  each element
$\widehat{O}_i^{\mu\nu}$ satisfies the gauge invariance condition
$q_{\mu}\widehat{O}_i^{\mu\nu}=0 $, and
so the corresponding form factors $F_i(p,q)$ are manifestly independent.

Writing
\bea
F_1(p,q)&=& -q^2 \sum_f\,\lambda_f \,G_a(x_f,y)\, ,
\nonumber \\
F_2(p,q)&=& -\sum_f\,\lambda_f\, G_b(x_f,y)\, ,
\nonumber\\
F_3(p,q)&=& 0 \, ,
\nonumber\\
F_4(p,q)&=& \sum_f\, \lambda_f\,G_c(x_f,y)\, ,
\nonumber\\
F_5(p,q)&=& 0 \, ,
\eea
we find
\bea
G_a( x,y) &=& 
 \frac{4}{y^3}
\left[\left( 2 - 3\,x + {x^3} \right) \,\Big( B_0(y) -  B_0(\bar{y})\Big)
+    {{\left( x -1  \right) }^3}\,\left( 2 + x \right) \,
       \left(C_0(x,y) + \frac{1}{x}C_0(\bar{x},\bar{y})\right)
\right.
\nonumber\\
&+&\left.
      {{\left(x -1  \right) }^2}\,\left( 2 + x \right) \,\log (x)
\right]
+ \frac{1}{y^2}\left[
\left( 1 - x \right) \,\left( 2 + x \right)
          - \frac{1}{3}\left( 42 - 23\,x + 5\,{x^2} \right) \,B_0(\bar{y})
\right.
\nonumber\\
&+& \left.
      \frac{4}{3}\,\left( 7 - 4\,x + 3\,{x^2} \right) \,
          B_0(y) + 
      \left(x -1 \right) \,\left( 18 -7\, x +x^2\right) \,
       C_0(x,y) 
\right.
\nonumber\\
&+&\left. 
      \frac{2}{x}\,\left(x -1  \right) \,
          \left( 4 -x +3\,x^2\right) \,
          C_0(\bar{x},\bar{y}) + 
      \left( 14 -3\, x + 4\,x^2\right) \,\log (x)
\right]
\nonumber\\
& +&
\frac{1}{y}\left[
{\frac{ \left( 1 + x \right) \,\left( 2 + x \right) }
        {4\,\left( 1 - x \right) }} + 
      \frac{1}{6}\left(x -34  \right) \,B_0(\bar{y})
         + \frac{1}{3}\left( 8 + x \right) \,
          B_0(y) -
      \left( 12 -6\, x +x^2\right)C_0(x,y) 
\right.
\nonumber\\
&+&\left. 
      2\,x\,C_0(\bar{x},\bar{y}) + 
      {\frac{\left( 34  -45\, x +18\,x^2+2\,x^3\right)
\,\log (x)}{6\,{{\left(x -1 \right) }^2}}}\right]
-2\,C_0(x,y)
\label{Ga}
\\
\nonumber \\
G_b( x,y) &=& 
\frac{4}{y^3}\left[
\left( 2 - 3\,x + {x^3} \right) 
\Big(B_0(\bar{y}) - B_0(y)\Big) -
{\left(x- 1 \right) }^3\,\left( 2 + x \right)
\Big( C_0(x,y) +\frac{1}{x} C_0(\bar{x},\bar{y})\Big)
\right.
\nonumber\\
&-&\left.
      {{\left( x -1  \right) }^2}\,\left( 2 + x \right) \,\log (x)\right]
 + \frac{2}{y^2}\left[
\left(x -1\right) \,\left( 2 + x \right)  
+  \frac{4}{3}\,\left( 3  -x + x^2 \right) \,
          B_0(\bar{y}) 
\right.
\nonumber\\
&-& \left.
      \frac{1}{3}\left( 2 + x + 9\,{x^2} \right) \,
          B_0(y) 
+     \left( 6 - 7\,x + 2\,{x^2} - {x^3} \right) \,
       C_0(x,y) 
\right.
\nonumber\\
&-&\left.   
  2\,\left(x -1 \right) \,\left( 1 + 2\,x \right) \,
       C_0(\bar{x},\bar{y}) 
- \left( 4 +  2\, x + 3\,x^2  \right) \,\log (x)\right]
 + 
\frac{1}{y}\left[
{\frac{3\,\left( 2 - 5\,x + {x^2} \right) }
        {2\,\left(x -1  \right) }} 
\right.
\nonumber\\
&+& 
\left.
      \frac{1}{3}\left( 2 - 5\,x \right) \,B_0(\bar{y})
         + \frac{2}{3}\,\left( 8 - 5\,x \right) \, B_0(y) + 
      \left( 4 - 2\,x \right) \,C_0(x,y) 
\right.
\nonumber\\
&-& \left. 
      \frac{1}{x}\left(4 - 6\, x  + 6\,x^2 \right) \,
       C_0(\bar{x},\bar{y}) 
-     {\frac{\left( 2 + 3\, x-24\, x^2 +10\, x^3                
\right) \,\log (x)}{3\,{{\left(x -1 \right) }^2}}}
\right]
-2\,C_0(\bar{x},\bar{y})
\label{Gb}
\\
\nonumber \\
G_c( x,y) &=& 
\frac{20}{y^3}\left[
  \left( 2 - 3\,x + {x^3} \right) \,\Big(
       B_0(\bar{y}) - B_0(y)\Big) - 
      {{\left(x -1 \right) }^3}\,\left( 2 + x \right) \, \Big(
       C_0(x,y) +           \frac{1}{x}C_0(\bar{x},\bar{y}) \Big)
\right.
\nonumber\\
&-&\left. 
      {{\left(x -1  \right) }^2}\,\left( 2 + x \right) \,\log (x)
\right] 
+ \frac{2}{y^2}\left[
8\,\left(x -1  \right) \,\left( 2 + x \right)  + 
      \frac{1}{3}\,\left( 78 -59\, x + 5\,x^2  \right) \,
          B_0(\bar{y}) 
\right.
\nonumber\\
&-& \left.
      \frac{1}{3}\,\left( 10 - 25\,x + 39\,{x^2} \right) \,
          B_0(y)
+ 18\,\left(x -2 \right) \,\left(x -1  \right) \,
       C_0(x,y) - 
      18\,x\,\left(x -1  \right)
       C_0(\bar{x},\bar{y}) 
\right.
\nonumber\\
&-&\left. 
      \left( 26 + 3\, x + 13\,x^2  \right) \,\log (x)
\right] + 
\frac{1}{y}\left[
{\frac{3\,\left( 2 - 5\,x + {x^2} \right) }{x -1 }} +
      {\frac{4\,\left(10 -x \right) \,
          B_0(\bar{y})}{3}} 
\right.
\nonumber\\
&+&\left. 
      \frac{4}{3} \left( 2 - 5\,x \right) \,
          B_0(y) + 
      2\,\left( 9 - x \right) \,\left( 2 - x \right)\, C_0(x,y)\, 
-  \frac{1}{x}\left( 4  - 14\, x  + 18\,x^2\right) \,
       C_0(\bar{x},\bar{y}) 
\right.
\nonumber\\
&-&\left. 
      {\frac{2\,\left( 20 -24\, x 
               -15\, x^2 + 10\,x^3 \right) \,\log (x)}
          {3\,{{\left(x -1 \right) }^2}}}\right]
+  4\, C_0(x,y) -2\, C_0(\bar{x},\bar{y})~,
\label{Gc}
\eea
\\ \noindent
where  $x_f\equiv m_f^2/m_W^2$ and $y\equiv q^2/m_W^2$, and
$\bar{x}\equiv 1/x$, $\bar{y}\equiv y/x$, and 
the functions $ B_0(y)$, $ C_0(x,y)$ are defined as
\bea
{B_0}(y)&=&
\frac{2}{y}\left(y-\sqrt{y\left(4-y\right)}\,
\arctan{\left[\frac{y}{\sqrt{y\left(4-y\right)}}\right]}\right)
\nonumber \\
{C_0}(x,y)&=&
\int_0^1\, dx_1\, \int_0^{1-x_1}\, dx_2\,\Big[(1-x)(x_1+x_2)+y\, x_1x_2-1
\Big]^{-1}\, .
\eea
An analytic result for the function ${ C_0(x,y)}$ can be obtained from
Refs.\cite{C0,rd}.
The functions $G_{a,b,c}({ x,y})$ generalize the $g_{a,b}({ x})$ 
in Eq.(\ref{functions}) to include the full $q^2$ dependence 
and satisfy the following conditions
\bea
{
\lim_{y\to 0} {\it G}_a(x,y)}=g_a( x)\, ,~~~~
{ \lim_{y\to 0} {\it G_b}(x,y)}={\it g_b}( x)\, ,~~~~
{ \lim_{y\to 0}{\it G_c}(x,y)}=0\, .
\eea
Notice that, despite the presence of pure $\log({ x})$ terms 
(not multiplied by $x$) 
in Eqs.(\ref{Ga})--(\ref{Gc}), the $G_{a.b.c}({ x,y})$ 
functions turn out to be of order ${\cal O}({ x})$ for small $x$.
This is because the $\log({ x})$ terms 
cancel out when summed to the corresponding terms proportional to the 
$ B(\bar{y})$ function. Indeed,
$ B(\bar{y})=\log(x) + {\cal O}(x)$. Therefore, as expected by 
the infrared behavior of the  diagrams in Fig.~1, 
for small internal fermion masses, 
the GIM mechanism strongly suppresses the TFCNC in all ranges of $q^2$.


\begin{thebibliography}{99}

\bibitem{Delb}
  R.~Delbourgo and P.~Phocas-Cosmetatos,
  Lett.\ Nuovo Cim.\  {\bf 5}, 420 (1972);
  Phys.\ Lett.\  B {\bf 41}, 533 (1972).

\bibitem{Grisaru}
  M.~T.~Grisaru, P.~van Nieuwenhuizen and C.~C.~Wu,
  Phys.\ Rev.\  D {\bf 12}, 1813 (1975).

\bibitem{BG1}
F.A. Berends and R. Gastmans, in {\it Proceedings of the Fifth International 
Conference on Neutrino Science}, edited by
A. Frenkel and G. Marx 
(Central Research Institute for Physics, Budapest, Hungary, 1975), 
p.310.

\bibitem{kob-okun}
  I.~Y.~Kobzarev and L.~B.~Okun,
  Zh.\ Eksp.\ Teor.\ Fiz.\  {\bf 43} (1962) 1904
  [Sov.\ Phys.\ JETP {\bf 16} (1963) 1343].

\bibitem{milton}
  K.~A.~Milton,
  Phys.\ Rev.\  D {\bf 15} (1977) 2149;
  Phys.\ Rev.\  D {\bf 15}, 538 (1977).


\bibitem{BG2}
  F.~A.~Berends and R.~Gastmans,
  Annals Phys.\  {\bf 98}, 225 (1976).


\bibitem{DeWitt}
  C.~M.~DeWitt and B.~S.~DeWitt,
  Physics {\bf 1}, 3 (1964).


\bibitem{CPRS}
  R.~Contino, L.~Pilo, R.~Rattazzi and A.~Strumia,
  JHEP {\bf 0106}, 005 (2001)
  [arXiv:hep-ph/0103104].


\bibitem{hlz}
  T.~Han, J.~D.~Lykken and R.~J.~Zhang,
  Phys.\ Rev.\  D {\bf 59}, 105006 (1999)
  [arXiv:hep-ph/9811350].

\bibitem{np}
  J.~F.~Nieves and P.~B.~Pal,
  Phys.\ Rev.\  D {\bf 72} (2005) 093006
  [arXiv:hep-ph/0509321].

\bibitem{grw}
  G.~F.~Giudice, R.~Rattazzi and J.~D.~Wells,
  Nucl.\ Phys.\  B {\bf 544}, 3 (1999)
  [arXiv:hep-ph/9811291].

\bibitem{bt}
  D.~Binosi and L.~Theussl,
  Comput.\ Phys.\ Commun.\  {\bf 161} (2004) 76
  [arXiv:hep-ph/0309015].

\bibitem{vDV}
  H.~van Dam and M.~J.~G.~Veltman,
  Nucl.\ Phys.\  B {\bf 22} (1970) 397.

\bibitem{zak}
  V.~I.~Zakharov,
  JETP Lett.\  {\bf 12} (1970) 312
  [Pisma Zh.\ Eksp.\ Teor.\ Fiz.\  {\bf 12} (1970) 447].

\bibitem{vainsh}
  A.~I.~Vainshtein,
  Phys.\ Lett.\  B {\bf 39}, 393 (1972).

\bibitem{ddgv}
  C.~Deffayet, G.~R.~Dvali, G.~Gabadadze and A.~I.~Vainshtein,
  Phys.\ Rev.\  D {\bf 65}, 044026 (2002)
  [arXiv:hep-th/0106001];
  G.~Dvali,
arXiv:hep-th/0402130.

\bibitem{ADD}
  N.~Arkani-Hamed, S.~Dimopoulos and G.~R.~Dvali,
  Phys.\ Lett.\  B {\bf 429}, 263 (1998)
  [arXiv:hep-ph/9803315].

\bibitem{AADD}
  I.~Antoniadis, N.~Arkani-Hamed, S.~Dimopoulos and G.~R.~Dvali,
  Phys.\ Lett.\  B {\bf 436}, 257 (1998)
  [arXiv:hep-ph/9804398].


\bibitem{rd}
  M.~Roth and A.~Denner,
  Nucl.\ Phys.\  B {\bf 479}, 495 (1996)
  [arXiv:hep-ph/9605420].

\bibitem{C0}
  G.~'t Hooft and M.~J.~G.~Veltman,
  Nucl.\ Phys.\  B {\bf 153}, 365 (1979);
  G.~J.~van Oldenborgh and J.~A.~M.~Vermaseren,
  Z.\ Phys.\  C {\bf 46}, 425 (1990).




\end{thebibliography}
\end{document}